\documentclass[preprint,12pt]{elsarticle}

\usepackage{graphicx}
\usepackage{amssymb}

\biboptions{numbers,sort&compress}

\usepackage{longtable}
\usepackage{url}
\newcommand{\tabincell}[2]{\begin{tabular}[t]{@{}#1@{}}#2\end{tabular}}

\journal{None}

\begin{document}

\begin{frontmatter}


\title{Brief View and Analysis to Latest Android Security Issues and Approaches}



\author{Ruicong Huang}

\address{Guangzhou, China}

\begin{abstract}
Due to the continuous improvement of performance and functions, Android remains the most popular operating system on mobile phone today. However, various malicious applications bring great threats to the system. Over the past few years, significant changes occured in both malwares and counter measures. Specifically, malwares are continuously evolving, and advanced approaches are adopted for more accurate detection. To keep up with the latest situation, in this paper, we conduct a wide range of analysis, including latest malwares, Android security features, and approaches. We also provide some finding when we are gathering information and carrying on experiments, which we think is useful for further researches and has not been mentioned in previous works.
\end{abstract}


\end{frontmatter}


\section{Introduction}
\label{S:1}

Android has become the most popular operation system on mobile phone for more than ten years. Nowadays, Android applications have covered the range of communication, business, education, economy, entertainment, etc. Due to the open-source feature, Android application developers have gained a lot of convenience, which also brings benefits to end users. However, these convenience also provides a good environment for malware developers. As more valuable information is stored on Android devices, the threat of malicious apps remains a concern for many users, because these malwares could cause privacy leaks or even economic loss. According to a report published by 360 Internet Security Center\cite{360lab2019report}, approximately 920 thousand new malicious program samples on Android were intercepted in the first half of 2019 (50 thousand every day on average). Among these malwares, privacy leak is the main type, accounting for 66.2\%, followed by expense cost (23.6\%), rogue behavior (4.9\%), remote control (4.3\%), malicious deduction (0.7\%) and fraud apps (0.2\%).

The fight against Android malwares never stops, and various approaches have been proposed to enhance the security of the system. In this case, it is necessary to have in-depth analyses to these researches, conclude pros and cons, and seek for improvements. As far as we know, existing works\cite{tam2017evolution, faruki2014android} have stayed at pre-2015 for a long time, and there is no further review to keep up with the latest researches. This paper aims at a comprehensive analysis towards the latest researches on Android security, most of which were proposed after 2015, and the methods they used to overcome the advanced penetration techniques. We collect the researches from three libraries: IEEE, ACM, and ScienceDirect. The contributions are listed as follows:

\begin{enumerate}
\item We present the latest situation of Android security (Section \ref{S:2}), including the evolution of Android malwares in last few years, and the security promotion of Android. Similar studies did not address this, but it is necessary for future work, because malwares are becoming more sophisticated, and knowledge of the malwares is of great help for counter measures. Besides, Android system have changed significantly with various security features, which could be combined with existing detection approaches.

\item We have a comprehensive collection and reasonable classification of the latest researches (Section \ref{S:3}). Unlike previous studies, we do not use common static/dynamic/hybrid categorization, because traditional static and dynamic analyses are known to have non-negligible flaws, and most works combines multiple features and methods. Instead, we classify them according to the approaches they used. We also have an analysis to their pros and cons.

\item We discuss some missing parts that are ignored by existing researches with detailed description or statistics, and introduce some promising directions for future works (Section \ref{S:4}).
\end{enumerate}

\section{Latest Situation}
\label{S:2}
Android has significant changes in the last few years with more security features, and provides a better experience for end users. However, the rapid evolution of malwares still brings great challenges. In this section, we first take a look at the security promotion of Android, and then describe the threats of novel malwares.

\subsection{Android Security Features}
The open-source feature of the Android system is one of the main reasons leading to the proliferation of Android malwares. Therefore, Google and Android developers have added security enhancement features to the system in version iterations. Android has five major version changes since 2015, from M (6.0) to Q Preview (10.0). Here, we discuss some major features.

\textbf{Dynamic permission granting mechanism (Android M, O)} Since Android M, permissions are no longer granted permanently like previous versions. They are divided into “normal” and “dangerous”, and each dangerous permission belongs to a specific permission group. When a dangerous permission is requested, the application is required to pop up a dialog box describing the related permission group to ask for granting consent. If the user agrees, all declared permissions in this group will be granted at the same time. In Android O, the mechanism is refined. Permissions in the same group as a granted permission should also be requested explicitly, or a “Permission Denial” exception will be thrown.

\textbf{Forbidding HTTP protocol (Android P)} Applications running on Android P are required to use HTTPS protocol for encrypted connections, and HTTP protocol is prohibited by default. If HTTP is truly needed in an application, the developer should make a specific declaration in the manifest file.

\textbf{Storage sandbox (Android Q Beta 1)} To help the users controlled the files at their will, the latest system uses new storage permission mechanism without “READ\_EXTERNAL\_STORAGE” and “WRITE\_EXTERNAL\_STORAGE”, and create storage sandbox for each application. With the sandbox, an application could access its own files on external storage without any permissions, and prevent unauthorized accesses from other applications. However, this mechanism may be too aggressive for most developers, and Google was forced to postpone it for the time being.

\textbf{Protection of unique device identifier (Android Q Beta 1)} Leak of unique device identifier, such as device id (IMEI), serial number and MAC address, is a long-standing problem. In Android P, Google rolled out a developer option to enable MAC address randomization, but it was turned off by default. With Android Q, the address will stay random even after you connect to a network, thus hiding your unique MAC. Besides, access for IMEI and serial number is restricted in Q, and the information is only available for platform and system apps. A privileged permission “READ\_PRIVILEGED\_PHONE\_STATE” is created and third-party apps cannot declare this permission.

\subsection{New Threats}
Android malwares has been evolving these years, and malwares of new families (as listed in Table \ref{tab:newMalwares}) keep occurring after 2015. Some of them successfully bypassed Bouncer, the official malware scanner of Google Play Store. With the development of virtual cryptocurrencies (bitcoin, monero, etc.), some malwares are developed to mine these coins for criminals in infected devices. Most of them are more sophisticated, compared to malwares in early stage, and more advanced techniques are used to hide themselves from detection. They do not show malicious behaviors at first. Instead, the malicious modules are activated using triggers, or by treating users to allow privileges. Most of them combine multiple abnormal behaviors.

\begin{longtable}{p{2.2cm}p{5cm}p{1.5cm}p{3cm}}
\hline
\textbf{Malware family (variants)} & \textbf{Risk and threats} & \textbf{First found in} & \textbf{First found by}\\
\hline
Gunpoder & Privacy leak & 2015 & Unit 42\cite{gunpoder}\\
~ & Adware\\
~ & Additional payload execution\\
Chamois & Botnet & 2016 & Google\cite{chamois}\\
~ & Premium SMS fraud\\
GM Bot & Obtaining root privilege & 2016 & IBM XForce\cite{gmbot1}\\
(Bankosy, & C\&C communication & ~ & FireEye\cite{gmbot2}\\
MazarBot, & Trojan\\
SlemBunk) & Ransomware\\
Triada & \tabincell{p{5cm}}{Zygote modification\\Remote access trojan\\Adware} & 2016 & Kaspersky Lab\cite{triada}\\
Ztorg & \tabincell{p{5cm}}{Trojan\\Premium rate SMS} & 2016 & FortiGuard Lab\cite{ztorg1, ztorg2}\\
Dvmap & Code injection & 2017 & Kaspersky\cite{dvmap}\\
Hiddad & Hidden ads & 2017 & Kaspersky Lab\cite{hiddad}\\
Xavier & \tabincell{p{5cm}}{Trojan\\Ad library} & 2017 & Trend Micro\cite{xavier}\\
GnatSpy & Privacy leak & 2017 & Trend Micro\cite{gnatspy}\\
Loapi & \tabincell{l}{Monero mining\\DDoS} & 2017 & Kaspersky Lab\cite{loapi}\\
FunkyBot & \tabincell{l}{C\&C communication\\Phishing} & 2018 & FortiGuard Lab\cite{funkybot}\\
Venus & Trojan & 2019 & GBHackers\cite{venus}\\
~ & Premium ads subscription\\
\hline
\caption{New Malwares on Android after 2015}
\label{tab:newMalwares}
\end{longtable}

\section{Approach Analysis}
\label{S:3}
In this section, we will first review traditional detection approaches briefly, and then introduce latest approaches in detail.
\subsection{Traditional Approaches}
There is something similar between malware detection on PC and on Android, since Android is an operating system based on Linux kernel, and Android apps are usually written in Java. Traditional methods can be broadly divided into static, dynamic, and hybrid analysis, and this classification is usually seen in existing reviews\cite{tam2017evolution, faruki2014android}. Static analysis analyzes the code without running the app. AndroidManifest.xml and classes.dex are the most common files to obtain the metadata, which contains permissions, components, call graph etc. Dynamic analysis depends on a virtual machine or sandbox, which is used to run the suspicious applications. The behaviors of the applications will be monitored and recorded for further analysis. Both methods have non-negligible limitations. Static analysis can only be adopted to analyze apps without code protection mechanisms, because it relies on a complete graph to find the paths between “sources” and “sinks”. Dynamic class loading, Java reflection and native methods could easily break the chains. Besides, it’s time-consuming and error-prone to analyze a whole set of paths. According to our experiment, using FlowDroid to find a taint propagation path from source “getDeviceId” method to sink “send http request” of an Android app “Today's Headline” (6.68MB with only two dex files) on a PC with Intel i5-6500 CPU and 8GB memory spends more than 10 minutes, because there are up to 70 thousand Java methods in this app. Out-of-memory exception is also frequently seen when constructing the graph. Dynamic analysis executes only one path per run, so the code coverage is not as good as static analysis. According to Wang's research\cite{wang2017netnlp}, performing dynamic analysis on mobile devices is challenging since it requires sufficient executions to improve code coverage. Some methods require modification to the source codes of Android systems, which increases difficulty of development and deployment. Besides, some advanced malwares will could detect mobile sandbox, such as Android AVD, genymotion, BlueStacks, and buildroid, which brings extra challenges to detection engines.

\subsection{Advanced Approaches}

\subsubsection{Graph-based Approaches}

Most of approaches of this kind aims at improvement of traditional static analysis by building a graph to describe the app's behaviors more accurately. The most representative research is Amandroid\cite{wei2014amandroid}. Amandroid made a large modification to the traditional static CG built by FlowDroid\cite{arzt2014flowdroid}. First, the inter-procedural control flow graph (ICFG), which contains edges representing inter-component communication (ICC), is constructed, so that control flow and data flow could pass through these edges like common function calls. Secondly, the data flow graph (DFG) and data dependency graph (DDG) are constructed based on ICFG to record the data pass, to perform various types of analysis, such as searching for specific data dependency chain or privacy leaks between specific sources and sinks. In April 2018, the authors provided an extended version\cite{wei2018amandroid}, including a new analysis algorithm and more experimental details.

Fan et al.\cite{fan2017dapasa} proposed DAPASA, an approach to detect repackaged apps using sensitive subgraph (SSG) analysis. An SSG is a subgraph of CG with the highest sensitivity, representing the most sensitive Android API. Two assumptions were established and TF-IDF model was adopted to calculate the sensitivity, which reflects the maliciousness of the application.

Monet proposed by Sun et al.\cite{sun2016monet} combined the runtime behavior graph (RBG) and the suspicious system call set to detect malware variants. Graph decoupling method is also used to improve accuracy and efficiency. Monet is a client-server structure. The client is installed in Android devices, to monitor malware behaviors and generate signatures by Binder call and system call interception, and the back-end server is used to evaluate the signature.

Feng et al.\cite{feng2015linkdroid} proposed LinkDroid to monitor unregulated usage of private information. Dynamic linkability graph (DLG) is used to track app-level linkability during runtime. Two apps are defined to be linkable if there is a path between them in DLG. The authors listed two kinds of sources of linkability that an adversary can exploit: OS-level information and inter-process communication (IPC). According to these information, LinkDroid is able to monitor linkability across apps.

Sokolova et al.\cite{sokolova2017android} introduced a detection approach with graph-based permission patterns. Applications are grouped into categories according to functionality, and a graph representing normal permission requests is built as the permission pattern of each category. The pattern contains seven associated metrics, which are used as inputs for classification feature construction, to verify abnormal permission requests and measure the risk level of an application.

Researches mentioned in this section are listed in Table \ref{tab:graphBasedApproaches} with their graph information and aims. The graph type is static or dynamic, meaning that the graph is constructed during static or dynamic analysis.

\begin{longtable}{p{2.6cm}p{4.7cm}p{2cm}p{2.7cm}}
    \hline
    \textbf{Research} & \textbf{Graph Name} & \textbf{Graph Type} & \textbf{Aim}\\
    \hline
    \tabincell{p{2.5cm}}{Wei et al.\\(Amandroid)} & \tabincell{p{4.7cm}}{Inter-procedural control flow graph\\Data flow graph\\Data dependency graph} & Static & \tabincell{p{2.5cm}}{Analysis\\Privacy leak detection}\\
    \tabincell{p{2.5cm}}{Fan et al.\\(DAPASA)} & Sensitive subgraph & Static & Piggybacked app detection\\
    \tabincell{p{2.5cm}}{Sun et al.\\(Monet)} & Runtime behavior graph & Dynamic & Malware variant detection\\
    \tabincell{p{2.5cm}}{Feng et al.\\(LinkDroid)} & Dynamic linkability graph & Dynamic & Monitoring\\
    Sokolova et al. & Permission pattern & Static & \tabincell{l}{Analysis\\Risk assessment}\\
    \hline
    \caption{Graph-based Approaches}
    \label{tab:graphBasedApproaches}
\end{longtable}

\subsubsection{Machine-learning-based Approaches}

Machine learning techniques have been widely used in detection. Generally, machine learning algorithms are used to build a classification model. These approaches usually focus on binary classification (verify whether the app is malicious or benign) or multiple classification (verify whether the app belongs to a particular malware family).

\textbf{Approaches with Single-layer Features}

System call and network data are two common sources that provide useful features for machine learning. The two types of research mainly focus on system call frequency/sequence pattern and HTTP packet, respectively.

Chen Da et al.\cite{da2016detection} proposed a malware detection method based on frequency analysis of system calls. App samples are divided according to their categories, such as tools, social networking apps, games, etc., to refine and distinguish various types of features. Then, a normalization method is used to process system call frequency information to improve detection accuracy. Finally, a random forest model is built to establish the training model.

Canfora et al.\cite{canfora2015detecting} focused on system call sequences, to learn the associations between the sequences and malicious behaviors. To obtain the most useful features, different length of subsequence is chosen with a feature selection step based on class difference. SVM with Gaussian radial kernel is used to train the classification model. In addition to dataset evaluation, the authors also assess their method in zero-day attacks using unseen execution trace, new malwares and new families.

Dimjašević, M. et al.\cite{dimjavsevic2016evaluation} considered both frequency and dependency representation, and implemented MALINE, an open-source tool. Based on the conclusion that a program's behavior can be characterized by dependencies formed through information flow between system calls, distances between system calls in the system call dependency graph are extracted as feature vectors, as well as the frequency of each call. The authors experimented several classification algorithms, including random forest, SVM, LASSO and ridge regression, and discussed some questions about validity, including hidden malicious behavior, architecture and randomness.

Ren et al.\cite{ren2016recon} proposed Recon, a real-time monitoring system for personally identifiable information (PII) leak based on network traffic tracking. More than 70 thousand data packets generated by applications were used as training samples, and crowdsourcing method was applied to guarantee the rationality of the data transmission. The author showed that a simple classifier based on C4.5 decision tree is able to identify PII leaks accurately, and is more efficient than ensemble methods.

Wang et al.\cite{wang2017netnlp} also focused on network traffic analysis. All HTTP packets are considered as plaintext documents, so that text semantics of network flows could be obtained using natural language processing (NLP) algorithm. Then, the authors proposed an automatic feature selection algorithm based on chi-square test to identify meaningful features for each HTTP document, and these features were used to build an SVM model for detection.

MalDozer proposed by Karbab et al.\cite{karbab2018maldozer} extracts features automatically using deep learning, to simplify the preprocessing step. API sequences are used as raw data to generate semantic vectors. As neural network could be applied to NLP problems\cite{kim2014convolutional}, apps, basic blocks, and APIs are treated as paragraphs, sentences, and words respectively. The system achieves good accuracy on datasets, and is easy to deployed, even on resource-limited devices such as Raspberry Pi. Similar work is Pektas et al.\cite{pektas2019learning}, using opcode sequences as raw data and a network of different architecture.

\textbf{Approaches with Multi-layer Features}

Andrea Saracino et al.\cite{saracino2018madam} proposed MADAM, a novel multi-level and behavior-based malware detector. 14 features extracted from four different levels (kernel, application, user, package) were used. Since all features are continuous, and it is assumed that the behavior of the malicious application have a large deviation from the standard, k-nearest neighbor (k-NN) algorithm, which is similarity-based, was chosen for classification. As for implementation, MADAM performs risk assessment of application metadata (before installation), system global monitoring and application behavior pattern recognition (at runtime). If malicious behavior occurs, a warning window is popped up to inform users.

DroidScribe proposed by Dash et al.\cite{dash2016droidscribe} focuses on dynamic analysis of runtime behavior, and is able to classify malware samples into families. The features used in DroidScribe for classification inherited the ones in CopperDroid\cite{tam2015copperdroid}, including files, network, binder process, and executes. To overcome the shortcomings of low coverage, the authors combined support vector machine (SVM) with conformal prediction to increase the accuracy of prediction, even in the case of insufficient information.

Papadopoulos et al. \cite{papadopoulos2017android} proposed a novel approach. Similar with DroidScribe, conformal prediction is also applied in this approach with a random forest classifier, to quantify uncertainty and provide provable confidence guarantees. Recorded features include six categories: battery, binder, CPU and memory usage, network, and permissions, and the values are calculated in six different ways.

\textbf{Adversarial Learning}

Adversarial learning is a new aspect, and it brings challenges to machine learning approaches. According to Grosse's research\cite{grosse2017adversarial}, most machine learning models are not robust enough against adversarial examples. They trained neural network for malware detection on Drebin dataset\cite{arp2014drebin}, and proposed a crafting algorithm to generate adversarial examples, which caused a misclassification rate of about 63\%. Further, they introduced two defense mechanisms (distillation\cite{papernot2016distillation} and adversarial training\cite{goodfellow2014explaining, szegedy2013intriguing}) to solve the problem.

Similarly, Chen X et al.\cite{chen2019androidhiv} proposed an innovative method of crafting adversarial examples to evade detection. The authors customized Grosse's algorithm and demonstrated a successful black-box attack on the original Drebin system. They also pointed that the perturbations can be implemented directly onto APK's Dalvik bytecode, rather than the Android manifest file only. Besides Drebin, the adversarial examples were also adopted against MaMaDroid with high evasion rates. Finally, they presented the effect of ensemble learning methods against the attack.

Chen S et al.\cite{chen2018kuafudet} proposed a complete solution, namely KuafuDet, against generated adversarial examples. KuafuDet is a two-phase learning enhancing approach, including an offline training phase and an online detection phase. The two phases are designed to be interactive and adversarial-based with a camouflage detector to identify the false negatives, so that the resilience and robustness of the system could be reinforced.

Researches listed above are summarized in Table \ref{tab:machineLearningBasedApproaches}. 

{\fontsize{9}{11}\selectfont
\begin{longtable}{p{2.2cm}p{2cm}p{1.5cm}p{2.4cm}p{1.6cm}p{1.4cm}p{1.5cm}}
\hline
\textbf{Research} & \textbf{Feature Range} & \tabincell{p{1.7cm}}{\textbf{Period of Obtaining Features}} & \tabincell{p{2.4cm}}{\textbf{Algorithm}\\\textbf{(usage)}} & \textbf{Aim} & \textbf{Dataset} & \textbf{Metrics}\\
\hline
Chen Da et al. & System calls & Runtime & Random forest & Malware detection & \tabincell{l}{Google Play\\Contagio} & Accuracy\\
Canfora et al. & System calls & Runtime & SVM & Malware detection & \tabincell{l}{Google Play\\Drebin} & \tabincell{l}{Accuracy\\FNR\\FPR}\\
\tabincell{p{2.2cm}}{Dimjašević, M. et al.\\(MALINE)} & System calls & Runtime & \tabincell{l}{Random forest\\SVM\\LASSO\\Ridge regression} & Malware detection & \tabincell{l}{Google Play\\Drebin} & \tabincell{l}{Accuracy\\Precision\\Sensitivity\\Specificity}\\
\tabincell{p{2.2cm}}{Ren et al.\\(Recon)} & Network traffic & Runtime & \tabincell{p{2.6cm}}{
Bag of words (feature extraction)\\
TF-IDF (feature selection)\\
C4.5 (model training)} & Leak detection & \tabincell{l}{Google Play\\AppsApk} & \tabincell{l}{Accuracy\\AUC\\FNR\\FPR}\\
Wang et al. & Network traffic & Runtime & \tabincell{p{2.3cm}}{NLP (preprocessing)\\
Chi-square test (feature extraction)\\
SVM (model training)} & Malware detection & \tabincell{l}{VirusShare\\Baidu} & \tabincell{l}{F1-Score\\FPR\\Precision\\Recall}\\
\tabincell{p{2.2cm}}{Andrea Saracino et al.\\(MADAM)} & \tabincell{l}{App metadata\\
APIs\\User activity\\System calls} & \tabincell{p{1.6cm}}{Installation (app metadata)\\Runtime (others)} &
k-NN & Real-time malware detection & \tabincell{l}{Genome\\Contagio\\VirusShare} & FPR\\
\tabincell{p{2.2cm}}{Dash et al.\\(DroidScribe)} & \tabincell{l}{Files\\Network\\Binder\\Executes} & Runtime & \tabincell{p{1.6cm}}{SVM with conformal prediction} & Family classification & Drebin & \tabincell{l}{Accuracy\\Precision\\Recall}\\
Grosse et al. & \tabincell{p{2cm}}{Permissions\\APIs\\Components\\Intents\\Network addresses} & Static analysis & DNN & Adversarial learning & Drebin & \tabincell{l}{FNR(MR)\\FPR}\\
Chen X et al. & \tabincell{p{2cm}}{Dexcodes\\Strings\\APIs\\Permissions} & Static analysis & \tabincell{p{2.6cm}}{Random forest\\k-NN (k=3)\\SVM\\DNN} & Adversarial learning & \tabincell{l}{PlayDrone\\Drebin\\VirusShare\\APKPure} & \tabincell{l}{F1-Score\\Evasion Rate}\\
\tabincell{p{2.2cm}}{Chen S et al.\\(KuafuDet)} & \tabincell{p{2cm}}{Permissions\\Intents and events} & Static analysis & \tabincell{p{2.6cm}}{SVM\\Random forest\\k-NN} & Adversarial learning defenses & \tabincell{l}{Google Play\\AMD\\Contagio\\Drebin} & \tabincell{l}{Accuracy\\FNR}\\
\hline
\caption{Machine-learning-based Approaches}
\label{tab:machineLearningBasedApproaches}
\end{longtable}
}

\subsubsection{Approaches for Native Code Analysis}

Native code is not an Android-specific feature, but programming code that is configured to run on a specific processor. For Android applications, Android native development kit (NDK) provides support for native development in C and C++ by using Java native interface (JNI), and the C/C++ code will be compiled into dynamic loading libraries. Most researches have limited analysis to native code, because it is low-level and is not architecture-independent like Dalvik bytecode. The convenience of native development and the difficulty of detection have brought a problem: native code has become a good choice for malwares to evade detection. To fill this gap, some approaches have been proposed.

To transfer instructions of different platforms (X86, ARM and MIPS) into unified patterns, Alam, S. et al. proposed MAIL\cite{alam2013mail}, an intermediate language. With MAIL, bytecode and native code of an application is transferred to a set of patterns for further analysis, such as DroidNative\cite{alam2017droidnative}, a malware detection system based on MAIL, annotated control flow graph and sliding window of difference, and DroidClone\cite{alam2016droidclone}, an approach to find code clones by comparing similarity between two MAIL programs.

Kalysch, A. et al.\cite{kalysch2018tackling} proposed novel improvements to the centroid approach, and implemented a plugin for IDA Pro disassembler for code similarity measures. The centroid approach works on a 3D-CFG, a 3-dimentional vector transferred from traditional CFG where each basic block has a unique coordinate. Native libraries are represented as the centroid, and parallelized DBSCAN is adopted for fingerprint-based parallel clustering. In addition to the accuracy, the authors also discussed the efficiency and the robustness against obfuscation.

Nowadays, Obfuscator-LLVM (O-LLVM)\cite{ollvm}, a project founded by Security Lab of Northwestern University of Applied Sciences in June 2010, is often used to protect native libraries from reverse engineering\cite{wong2018tackling, lim2017anti}. At the same time, deobfuscation techniques\cite{kan2019automated, francis2014deobfuscation} is also promising, although there are not much analyses for the time being.

In general, native code analysis is a challenging topic of Android security, and relies heavily on manual analysis with disassembler like IDA Pro\cite{idapro}. However, this issue could be handled from other aspects, because regardless of whether the program is written in bytecode or native code, the process will be under supervision of Android built-in security mechanisms (permission mechanism, Linux UGO model, SELinux, etc.). To successfully launch an attack, the malware containing native code must be accompanied by other sensitive behaviors, such as obtaining root privilege or sending SMS, which are more obvious and easier to detect. Existed researches and reports\cite{gunpoder, chamois, gmbot1, gmbot2, triada, ztorg1, ztorg2, xavier, loapi, funkybot, venus, currentandroidmalware} have confirmed that most malwares have more than one potentially harmful behavior.

\section{Discussion}
\label{S:4}
\subsection{Finding}
\subsubsection{Abandoned Android Malwares}
\begin{figure}[h]
\centering
\includegraphics[width=9cm]{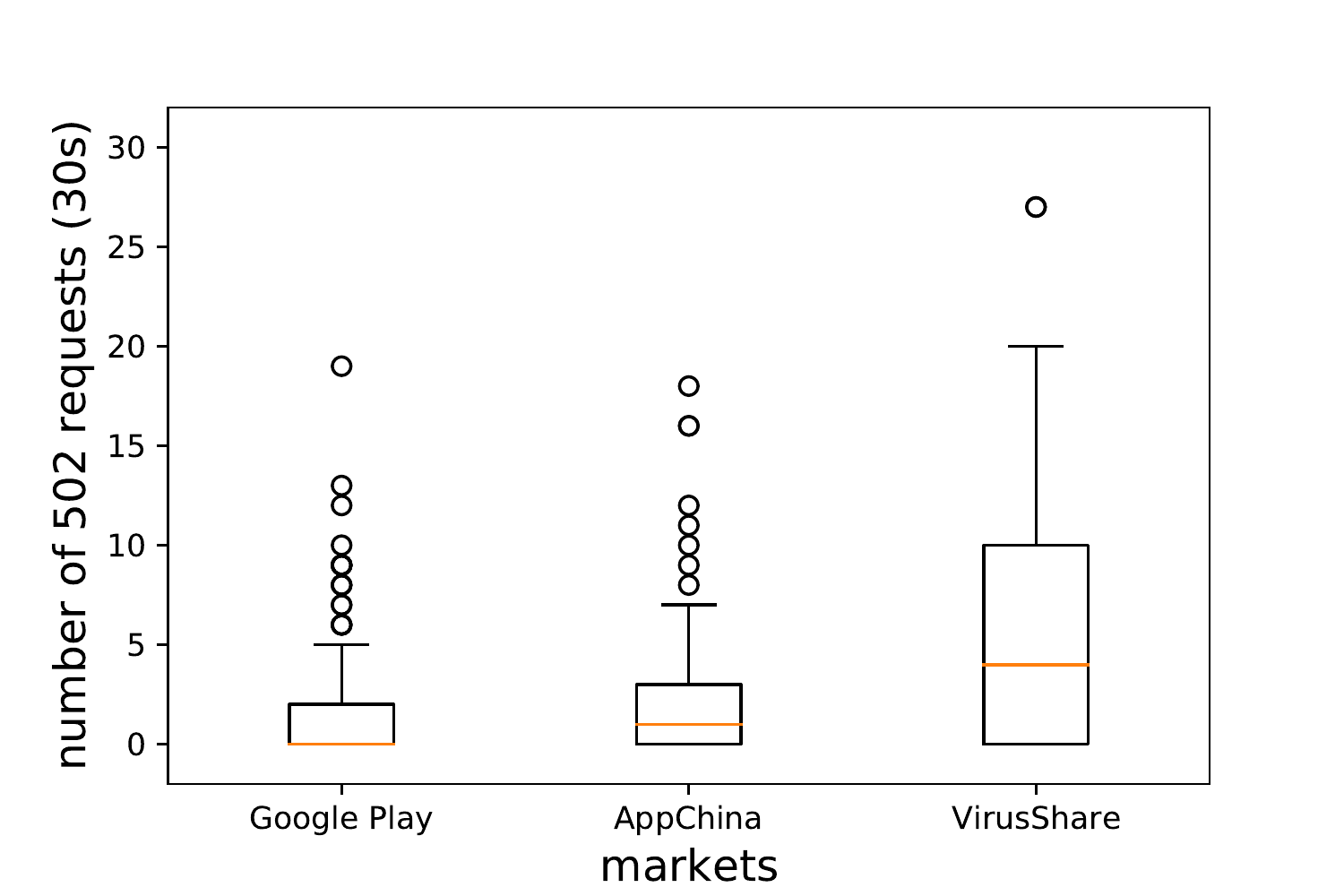}
\caption{Number of HTTP502s of apps from different markets}
\label{fig502}
\end{figure}

With the development of new malwares, a lot of old malwares are abandoned at the same time. We found an interesting phenomenon when we are conducting traffic analysis on Android malwares: there are a lot of HTTP requests with 502 response code (hereinafter referred to as HTTP502). 502 response code means a bad gateway, and it is a common server-side error when the address of the server could not be found. In our opinion, these servers are either closed by the cloud manager, if they are found to be engaged in illegal activities, or migrated by the malware developers, to better hide themselves from detection. Once the servers are migrated, malware developers will release a batch of new apps, and the old ones will be abandoned. When the old apps send requests to the original IPs or hosts, the server will always return 502 responses. However, the old apps may run without visible errors if exceptions are correctly handled. We collected 200 apps from Google Play, AppChina and VirusShare respectively, and ran each app on Xiaomi 3S for 30 seconds. Fig \ref{fig502} shows the distribution of HTTP502 number when the app is running.

The result shows that malwares usually have more HTTP502s than benign apps. 
Based on these results, we conclude that the number or frequency of requests with 502 response could be used as a feature to verify abandoned malwares. The same goes for evaluating the quality of Android applications of a market. Obviously, apps from Google Play have higher quality than AppChina, because most of them have less HTTP502s.

Abandoned malwares are frequently seen in datasets built in early stage, especially in drebin dataset. These malwares lose the ability to cause damages to the Android system and steal privacy for the time being, and their characteristics may be different from before (when they were active). The inactivity may be temporary or permanent, depending on whether the attacker restarts the server.

\subsubsection{FileObserver and Sensitive Files}
File access is an important characteristic, since it could provide useful features for further analyses. After inspection of researches using file access analysis, we find that most of them analyze file access events by monitoring specific Java APIs or system calls. The two methods are effective indeed, but still have shortcomings. Methods based on API monitoring have blind spots. For a simple example, external commands such as \texttt{Runtime.getRuntime().exec("cat <filename>")} could evade the detection, and more obfuscations could be developed. System calls like “open”, “close”, “read”, “write”, etc. could reflect file events, but strace with root privilege is needed to collect the calls. The lifecycle of strace process is difficult to handle automatically when analyzing multiple apps, because only one strace process is allowed at a time. Stracing zygote process is not a good idea either, because a modified Android system is needed, and the output is too redundant

In our opinion, the best way to monitor a file is to monitor the file itself. The Linux kernel has introduced a file change notification mechanism, Inotify, since version 2.6.13, which can efficiently track changes in the Linux file system in real time. Android is a linux-kernel-based operating system, and almost all versions of Android support Inotify. Furthermore, Android encapsulates a FileObserver class to facilitate the use of the Inotify mechanism. Using Java-implemented FileObserver is indeed more effective and convenient than monitoring APIs and system calls, since it is obfuscation-resistant and could be easily integrated into Android apps and Xposed plugins. To our knowledge, FileObserver is few mentioned in implementation of file access monitoring. We believe that the use of FileObserver could improve the robustness of file analysis on Android.

\begin{longtable}{p{4.4cm}p{4cm}p{4cm}}
\hline
\textbf{File} & \textbf{Associated privacy} & \textbf{Access}\\
\hline
\tabincell{l}{/data/data/com.android.\\
providers.telephony/dat\\
abases/mmssms.db} & Short messages and multimedia messages & Can read/write with root privilege\\
\tabincell{l}{/data/data/com.android.\\
providers.contacts/data\\
bases/contacts2.db} & Contact information & Can read/write with root privilege\\
\tabincell{l}{/data/data/com.android.\\
providers.telephony/dat\\
abases/telephony.db} & Sim card information & Can read/write with root privilege\\
\tabincell{l}{/sys/class/net/eth0/add\\ress} & MAC address & Can read without any permissions\\
\tabincell{l}{/sys/class/net/wlan0/ad\\dress} & WiFi MAC address & Can read without any permissions\\
/proc/cpuinfo & CPU usage information & Can read without any permissions\\
/proc/meminfo & Memory usage information & Can read without any permissions\\
\hline
\caption{Sensitive Files}
\label{tab:sensitiveFiles}
\end{longtable}

After analysis experiments of some malwares and the detection results of VirusTotal\cite{virustotal} (section “file system actions” of tab “behavior”), we list some sensitive files in Table \ref{tab:sensitiveFiles}. Direct access to these files is probably related to privacy leaks. With FileObserver, these files could be watched, even in rooted devices.

\subsubsection{Datasets}

To proof the effectiveness and good performance, most researchers will evaluate their approaches with Android app samples. If the selected dataset reflects the real situation, the proposed approach will probably achieve good results in real scenarios too. After summing up the dataset usage, we found that Drebin is the most frequently-used malware dataset. As for benign app samples, most researches chose Google Play, the official Android app store of Google, due to the strict investigation and guaranteed application quality. Other popular markets are less considered.

As mentioned in Tam's review, datasets should be updated regularly with newer samples for continuous evaluation of a system. However, most malware datasets are unable to keep up with the evolution. Drebin is a typical example. To fill this gap, we introduce some newer datasets constructed by senior researchers, security organizations or institutions. Some of them also contain benign samples. Update of benign samples does not need to be worried about, since the latest samples could be crawled from app markets at any time.

AndroZoo: AndroZoo is a growing collection of Android Applications founded by Allix et al. containing 10,468,689 apps. Both benign and malicious samples are included.

Koodous: Koodous is a collaborative platform with analysis tools and social interactions between the analysts. Users could vote and leave comments for app samples. The number of app samples is over 55 million and is still increasing, including more than 15 million detected samples and 250 thousand potential malwares.

Sk3ptre's GitHub: The author collected Android malware samples from 2018 to 2020, and built 3 repositories. Malwares of each family are compressed in a zip file. A Readme file is provided with SHA256 value of each app.

Ashishb's GitHub: There are 298 malware samples of 39 families in this project as of the writing date. Each family forms a folder with a Readme file recording the source.

\subsection{Future Work Prediction}

\subsubsection{Study on Packing Techniques}
\begin{figure}[h]
\centering
\includegraphics[width=10.5cm]{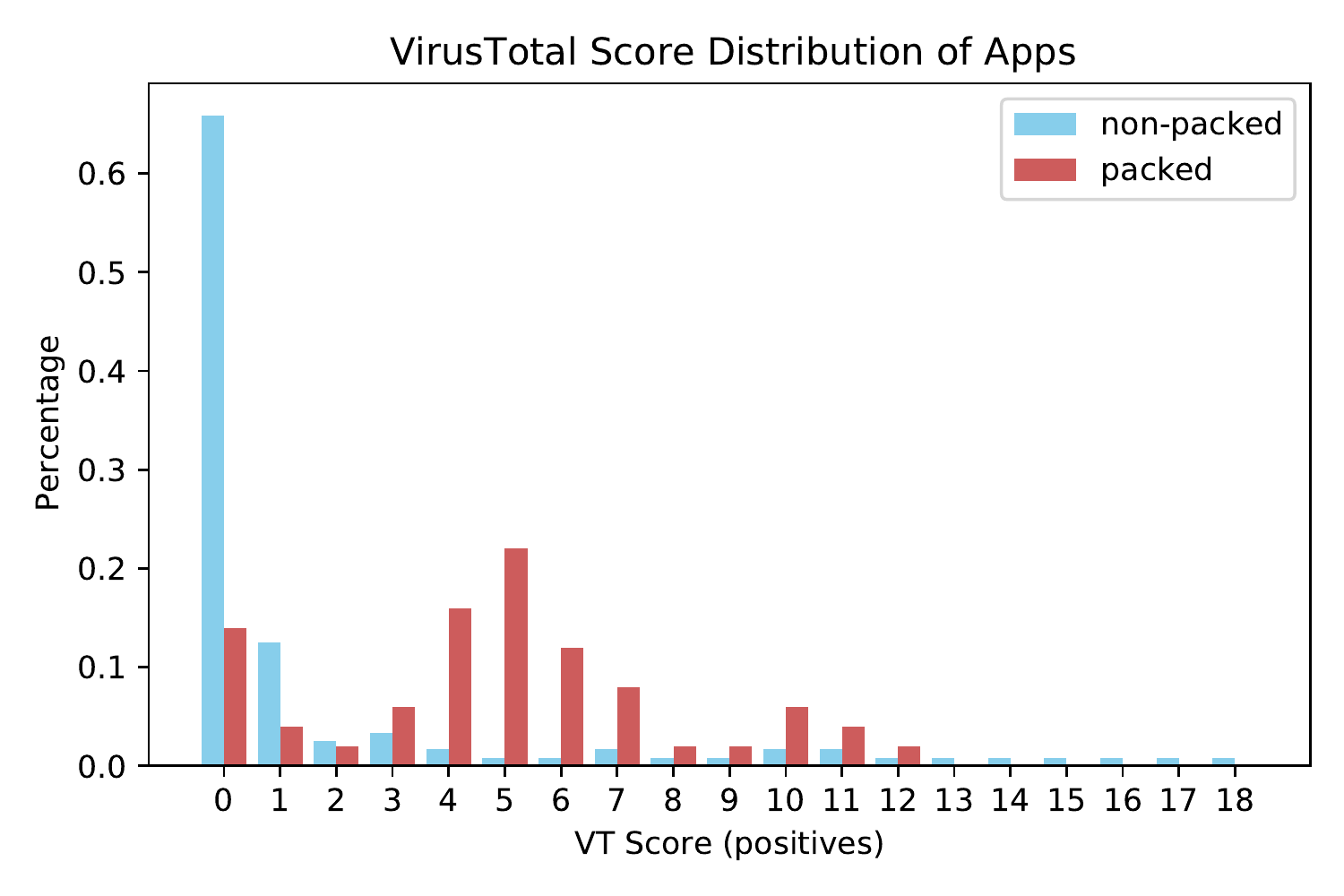}
\caption{VirusTotal score distribution of packed and non-packed apps}
\label{figPacking}
\end{figure}

Packing techniques are usually used to protect codes from decompiling. However, packing techniques updates frequently, and most of them have not been studied. Nowadays, many detection engines still consider packers as threats. We randomly selected 200 non-packed and 200 packed apps from AppChina, and uploaded them to VirusTotal. The score distribution is shown in Fig. \ref{figPacking}.

According to the results, we found that using packer, such as libjiagu (packer of Qihoo 360), libBaiduProtect (packer of Baidu) and ijiami, will greatly increase the rate of being detected as a malware in VirusTotal. The Pearson correlation coefficient between ‘found\_packers’ and VirusTotal score reaches 0.34, which is even higher than some sensitive permissions like SEND\_SMS (0.193) and SYSTEM\_ALERT\_WINDOW (0.292). The reason of this phenomenon may be entrypoint substitution, which means that the packers use their own Activities to replace the original entries when the app is “onCreated”. This behavior has something similar with app repackaging or Trojan, so this technique is rarely used in popular apps such as WeChat and Today's Headline. However, some detection engines may remain negative if they could recognize the packer. Packed apps are also rarely discussed in most researches, except those that specialize in (un)packing, such as AppSpear\cite{li2018appspear}, DexHunter\cite{zhang2015dexhunter}, PackerGrind\cite{xue2017adaptive}, and DexX\cite{sun2018dexx}.

\subsubsection{Solutions for More Android Device}
Internet of things (IoT) technology plays an important role in making things more intelligent. Nowadays, smart devices are not limited to mobile phones. A lot of smart home devices, watches, vending machines, vehicles and other new types of IoT devices are invented with rapid speed. Android-based systems (Wear OS, Android TV, Android Things, etc.) are widely adopted in these devices, and they face various threats too. Tam's review\cite{tam2017evolution} has pointed out this problem, but we want to add that it is not easy to find a universal solution for IoT devices, due to the heterogeneity of platforms and (altered) Android versions. Similarly, datasets for malware detection on mobile phones are no longer suitable, since these apps are not designed to run on IoT devices.

\section{Conclusion}
Android is facing real security problems today, so the malware detection should not be only talked about on papers. According to our research, although researches on Android security has covered many aspects, and made great progress, there are still room for approaches on production environment and evaluation of real-world apps. In this paper, we make extensive introductions and analyses towards the security issues and state-of-the-art researches. In addition, we provide information about abandoned malwares and implementation of a file access monitoring mechanism. Compared with previous reviews, this paper provides newer information and a different taxonomy of latest approaches.






\bibliographystyle{elsarticle-num-names}
\bibliography{sample.bib}







\end{document}